\documentclass[a4paper,11pt]{article}
\usepackage[top=1.8cm,bottom=1.8cm,left=1.8cm,right=1.8cm]{geometry}
\usepackage{xspace}
\usepackage{tablefootnote}

\newcommand{\lrs}{\emph{lrs}\xspace}
\newcommand{\lrsa}{\emph{lrs1}\xspace}
\newcommand{\plrs}{\emph{plrs}\xspace}
\newcommand{\mplrs}{\emph{mplrs}\xspace}
\newcommand{\mplrsa}{\emph{mplrs1}\xspace}
\newcommand{\mai}{\emph{mai12}\xspace}
\newcommand{\mait}{\emph{mai20}\xspace}
\newcommand{\maitf}{\emph{mai24}\xspace}
\newcommand{\mais}{\emph{mai64}\xspace}
\newcommand{\mainew}{\emph{mai32abcd}\xspace}
\newcommand{\mainewer}{\emph{mai32ef}\xspace}

\newcommand{\cdd}{\emph{cddr+}\xspace}
\newcommand{\porta}{\emph{porta}\xspace}
\newcommand{\ppl}{\emph{ppl}\xspace}

\newcommand{\tsubame}{\emph{Tsubame2}\xspace}
\newcommand{\norm}{\emph{normaliz}\xspace}
\newcommand{\normaliz}{\emph{normaliz}\xspace}

\usepackage{amsmath}
\usepackage{amssymb}
\usepackage{latexsym}
\usepackage{graphicx}
\usepackage{color}
\usepackage{hyperref}
\usepackage{slashbox}
\usepackage{algorithm}
\usepackage{algpseudocode}

\definecolor{darkblue}{rgb}{0,0,0.6}
\hypersetup{
	citecolor=darkblue,
	linkcolor=darkblue,
	urlcolor=darkblue,
	breaklinks=true,
	colorlinks=true,
	citebordercolor=0 0 0,
	filebordercolor=0 0 0,
	linkbordercolor=0 0 0,
	menubordercolor=0 0 0,
	urlbordercolor=0 0 0,
	pdfhighlight=/I,
	pdfborder=0 0 0,
	bookmarksopen=true,
	bookmarksnumbered=true
}

\include{./mathlib}
\include{./macros}

\begin{document}

\title{Comparative computational results for some vertex and facet enumeration codes}
\author{David Avis \\ School of Informatics, Kyoto University, Kyoto, Japan and
   \\ School of Computer Science,
    McGill University, Montr{\'e}al, Qu{\'e}bec, Canada
\and Charles Jordan \\
 Graduate School of Information Science and Technology,
Hokkaido University, Sapporo, Japan}
\maketitle
\begin{abstract}
We report some computational results comparing parallel and sequential codes
for vertex/facet enumeration problems for convex polyhedra. The problems chosen span the 
range from simple to highly degenerate polytopes. 
We tested one code (\lrs) based on pivoting and four codes (\cdd, \ppl, \norm,
\porta) based on the double description method.
\norm employs parallelization as do the codes \plrs and \mplrs which are based on \lrs.
We tested these codes using various hardware configurations with up to 1200 cores.
Major speedups were obtained by parallelization, particularly by the code \mplrs which 
uses MPI and can operate on clusters of 
machines.
\end{abstract}

\section{Background and polytopes tested}
\label{back}
A convex polyhedron $P$ can be represented by either a list of vertices and extreme rays,
called a V-representation, or a list of its facet defining inequalities, called
an H-representation. 
The vertex enumeration problem is to convert
an H-representation to a V-representation. 
The computationally equivalent facet
enumeration problem performs the reverse transformation. 
For further background see G. Ziegler \cite{Ziegler}. 

In this note we consider only polytopes (bounded polyhedra)
so extreme rays will not be required. Furthermore, for
technical simplicity in this description, we assume that all polytopes
are full dimensional. Neither condition is required for the algorithms
tested and in fact some of our test problems are not full dimensional. 
The input for either problem is represented by an $m$ by $n$ matrix.
For the vertex enumeration problem this is a list of $m$ inequalities 
in $n-1$ variables whose intersection define $P$. For a facet
enumeration problem it is a list of the vertices of $P$ each beginning
with a 1 in column one\footnote{Extreme rays would be indicated by a zero in column one.}.
So in either case,
under our assumption, the dimension of $P$ is $n-1$.

One of the features of this type of enumeration problem is that the output
size varies widely for given input 
parameters $m$ and $n$. This is shown
explicitly by McMullen's Upper Bound Theorem (see, e.g., \cite{Ziegler})
which is tight.
It states that for a vertex enumeration problem with parameters $m, n$ we have:
\def\lf{\left\lfloor}   
\def\rf{\right\rfloor}
\begin{equation}
|V| \le \binom{m- \lf \frac{n}{2} \rf }{m-n+1}+ \binom{m- \lf \frac{n+1}{2} \rf }{m-n+1}
\label{ubt}
\end{equation}
where $|V|$ is the number of vertices 
that are output.
For a facet enumeration problem, by polarity of polytopes, the
same inequality holds if we replace $|V|$ by $|F|$, the number of facets. 
By inverting the formula we can get lower bounds on the output size.

A class of polytopes for which the bound (\ref{ubt}) is tight are the $cyclic~polytopes$
which are usually given as a V-representation consisting of $m$ points on the 
$(n-1)$-dimensional moment curve.
So, for example, a cyclic polytope with $m=40$ and $n=21$ has $|V|=40$ vertices in dimension 20 and
$|F|=40,060,020$ facets. This implies that if we started with its H-representation,
i.e., $m=40,060,020$ and $n=21$, then the output would consist of only 40 vertices!
Problems of this second type are called {\em highly degenerate} since each vertex
may be described by many different combinations of facets. This contrasts
with a {\em simple} polytope where each vertex is given by the intersection of exactly
$n-1$ facets. Dually a {\em simplicial} polytope is one where each facet contains precisely
$n-1$ vertices. Cyclic polytopes are simplicial.

The polytopes we tested are described in Table \ref{polytopes} and range from simple polyhedra
to highly degenerate polyhedra. This table includes the results of an \lrs run on each
polytope as  \lrs
gives the number of cobases in a symbolic perturbation of the polytope,
showing how degenerate the polytope is.
The corresponding input files are contained in the lrslib-062 distribution~\cite{lrs} in subdirectory
\texttt{lrslib-062/ine/test-062}.
Note that the input sizes are small, roughly comparable and, except for $cp6$, much smaller than the output
sizes.
Five of the problems were previously used in \cite{AR13}:
\begin{itemize}
\item {\em c30-15}, {\em c40-21}\/: cyclic polytopes described above.
These have very large integer coefficients, the longest having 23 digits 
for {\em c30-15} and 33 digits for {\em c40-21}.
\item {\em mit}\/: a configuration polytope used in materials science, created by G.\ Garbulsky~\cite{ceder1994}.
The inequality coefficients are mostly integers in the range $\pm 100$ with a few larger values. 
\item {\em perm10}\/: the permutahedron for permutations of length $N=10$, whose vertices are
the $10!$ permutations of $(1,2,3,...,10)$. It is a 9-dimensional simple
polytope. More generally, for permutations of length $N$, this polytope is described by $2^N -2$ facets and one equation
and has $N!$ vertices.
The variables all have coefficients $0$ or $1$.
\item {\em bv7}\/: an extended formulation of the permutahedron based on the  Birkhoff-Von Neumann polytope.
It is described by $N^2$ inequalities and $3N-1$ equations in $N^2+N$ variables and also has $N!$ vertices.
The inequalities are all $0,\pm 1$ valued and the equations have single digit integers. The input matrix
is very sparse and the polytope is
highly degenerate.
\end{itemize}

The new problems are:
\begin{itemize}
\item {\em fq48-19}\/: related to the travelling salesman problem for $N=5$, created by F.\ Quondam
(private communication).
The coefficients are all $0,\pm 1$ valued and it is moderately degenerate.
\item {\em mit71}: a correlation polytope related to problem {\em mit}, created by G.\ Garbulsky~\cite{ceder1994}.
The coefficients are similar to {\em mit} and it is moderately degenerate.
\item {\em zfw91}\/: $0, \pm1$ polytope based on a sensor network that is extremely degenerate
and has large output size, created by Z.F.\ Wang~\cite{zfw14}. There are three non-zeroes per row.
\item {\em cp6}\/: the cut polytope for $N=6$ solved in the `reverse' direction: from an H-representation to a V-representation.
The output consists of the 32 cut vectors of $K_6$. It is extremely degenerate,  approaching the lower bound
of 19 vertices implied by (\ref{ubt}) for these parameters.
The coefficients of the variables are $0, \pm 1, \pm 2$.
\end{itemize}

\renewcommand*{\thefootnote}{\fnsymbol{footnote}}
\begin{table}[htbp]
\centering
\scalebox{0.9}{
\begin{tabular}[t]{|c||c|c|c|c||c|c||c|c|c|} 
  \hline
  Name &  \multicolumn{4}{|c||}{Input} & \multicolumn{2}{|c||}{Output} & \multicolumn{3}{|c|}{\lrs}\\
       & H/V & $m$    & $n$ &size    & V/H  & size  & bases &depth & secs \\
  \hline
{\em bv7} & H & 69 & 57 & 8.1K & 5040& 867K          &84707280  &17 &8300  \\
{\em c30-15} & V & 30 & 16 & 4.7K& 341088 & 73.8M &319770& 14 & 39 \\
{\em c40-20} & V & 40 & 21 & 12K& 40060020 & 15.6G &20030010   & 19 & 9445 \\
{\em fq48-19} & H & 48 & 19 & 1.3K& 119184 &8.7M    & 7843390  &24 & 251 \\
{\em mit71} & H & 71 & 61 &9.5K& 3149579 &1.1G     & 57613364 &20 & 20688\\
{\em mit} & H & 729 & 9 & 21K& 4862 & 196K         & 1375608 &101 & 496\\
{\em perm10} & H & 1023 & 11 & 29K& 3628800 &127M  & 3628800  &45 & 2230 \\
{\em zfw91} & H & 91 & 38 &7.1K & 2787415 &205M  & 10819289888124\tablefootnote{Computed by \mplrsa v6.2 in 2144809 seconds using 289 cores (see Section \ref{alg}).}  & & * \\
\hline
{\em cp6} & H & 368 & 16 &18K& 32& 1.6K &4844923002&153 &1762156\tablefootnote{Computed by \lrs v6.0}\\

  \hline
\end{tabular}
}
\caption{{Polytopes tested and \lrs times(\mait):  *=time $>$ 604800 secs}}
\label{polytopes}
\end{table}
\renewcommand*{\thefootnote}{\arabic{footnote}}

\section{Algorithms, implementations and machines used}
\label{alg}
There are basically
two approaches to this problem: pivoting using reverse search \cite{AF92} and the 
Fourier-Motzkin double description method, see \cite{Ziegler}. 
The conventional wisdom is to
use the double description method if the polytope is highly degenerate and use a pivoting
method if it is simple or has low degeneracy (see e.g.\ \cite{polymake}, Section 3).
Results below shed some doubt on the first part of this rule, especially when parallel processing is used.
They do, however, confirm the second part of the rule.
We tested five sequential codes, including four based on the
double description method and one based on pivoting:
\begin{itemize}
\item
\cdd(v.\ 0.77): Double description code developed by K.\ Fukuda~\cite{cdd}. 
\item
\normaliz(v.\ 3.0.0): Hybrid parallel double description code developed by the Normaliz project~\cite{norm}.
\item
\porta(v.\ 1.4.1): Double description code developed by T.\ Christof and A.\ Lobel~\cite{porta}.
\item
\ppl(v.\ 1.1): Double description code developed by the Parma Polyhedral Library project~\cite{ppl}.
\item
\lrs(v.\ 6.2): C vertex enumeration code based on reverse search developed by D.\ Avis~\cite{lrs}.
\end{itemize}
\noindent
Of these five codes, \lrs and \normaliz offer parallelization. For \normaliz this
occurs automatically with the standard implementation if it is run on a shared memory multicore machine.
The number of cores used can be controlled with the -x option, which we used extensively in our tests.
For \lrs two wrappers have been developed:
\begin{itemize}
\item
\plrs(v.\ 6.2): C++ wrapper for \lrs using the Boost library, developed by G.\ Roumanis~\cite{AR13}.
It runs on a single shared memory multicore machine.
\item
\mplrs(v.\ 6.2): C wrapper for \lrs using the MPI library, developed by the authors~\cite{AJ15b}. It runs on a network of multicore machines.
\end{itemize}
\noindent
For {\em cp6} the \lrs times in Tables \ref{polytopes}--\ref{tab:single}
and the \plrs times in Table \ref{tab:su64} 
were obtained using v.\ 6.0 which has a smaller backtrack cache size than v.\ 6.2.
Hence the \mplrs speedups against \lrs for {\em cp6} 
in Table \ref{tab:single} are probably somewhat larger than they would be against \lrs v.\ 6.2.
The \mplrs times in Table \ref{tab:su1200} were obtained using v.\ 5.1b.

All of the above codes compute in exact integer arithmetic and with the
exception of \porta, are compiled with the GMP library for this purpose.
However \normaliz uses hybrid arithmetic, giving a very large speedup for
certain inputs as described in the next section.
In addition, \porta can be run in either fixed or extended precision.

Finally, \lrs is also available in a fixed precision 64-bit version, \lrsa, which does no overflow checking. 
In general, this gives unpredictable results that need independent verification.
In practice, for cases when there is no arithmetic overflow,
\lrsa runs about 4--6 times faster than \lrs (see Computational Results on
the \lrs home page~\cite{lrs}).
The parallel version of \lrsa,
\mplrsa, was used to compute the number of cobases for {\em zfw91}, taking roughly 25 days on 289 cores.

The tests were performed using the following machines:
\begin{itemize}
\item
\mait: 2x Xeon E5-2690 (10-core 3.0GHz), 20 cores, 128GB memory, 3TB hard drive
\item
\mainew: 4 nodes, each containing: 2x Opteron 6376 (16-core 2.3GHz), 32GB memory, 500GB hard drive (128 cores in total)
\item
\mainewer: 4x Opteron 6376 (16-core 2.3GHz), 64 cores, 256GB memory, 4TB hard drive
\item
\mais: 4x Opteron 6272 (16-core 2.1GHz), 64 cores, 64GB memory, 500GB hard drive
\item
\mai: 2x Xeon X5650 (6-core 2.66GHz), 12 cores, 24GB memory, 60GB hard drive
\item
\maitf: 2x Opteron 6238 (12-core 2.6GHz), 24 cores, 16GB memory, 600GB RAID5 array
\item
\tsubame: supercomputer located at Tokyo Institute of Technology
\end{itemize}
The first six machines total 312 cores and are located at Kyoto University. They were purchased between 2011-15 for a combined
total of 3.9 million yen (\$33,200).

\section{Computational results}
Table \ref{tab:single} contains the results obtained by running the five sequential codes
on the problems described in Table \ref{polytopes}.
The times for \lrs shown in Table \ref{polytopes} are included for comparison.
The time limit was one week (604,800 seconds) except for {\em cp6}.
Programs \cdd, \lrs, \ppl were used with no parameters.

\renewcommand*{\thefootnote}{\fnsymbol{footnote}}
\begin{table}[htbp]
\centering
\scalebox{0.9}{
\begin{tabular}{|c||c||c||c||c|c||c|c|}
 \hline

Name     &{\lrs}& \cdd & \ppl&\multicolumn{2}{|c||}{\norm} &\multicolumn{2}{|c|}{\porta}   \\  
           &  secs   & secs  & secs& secs(hybrid) & secs(GMP) & secs(64-bit) & secs(extended)  \\
\hline
{\em bv7}    &  8300   &  *    & 578    & 122 &1030 &315& 310\\ 
{\em c30-15}     &  39   &  2991   &  3040     & ** &** & ** & **\\
{\em c40-20}     &  9445   &  *    &  *     & **  &** & ** & **\\
{\em fq48-19}  &   252    & 437   & 1355   &  41&300& 5103 & 4561 \\
{\em mit71}     &  20688  &  *    & 260347 & 503564 & 364354& 108993 & 107689\\
{\em mit}       &   496    & 368   & 40644  & 175 &2174 & ** & 47478\\
{\em perm10}    &  2230   &  *    &  *     &  1025&33240 & * &* \\
{\em zfw91}   &  *   &  *    &  *     &  189763&* & 31348 & 30787\\
\hline
{\em cp6}       & 1762156\tablefootnote{Computed by \lrs v6.0} &1463829&$>$6570000 & 138162& 1518785 &  **& $>$4925580 \\
\hline
\end{tabular}
}
\caption{{Single processor times(\mait):  *=time $>$ 604800 secs  **=abnormal termination }}
\label{tab:single}
\end{table}
\renewcommand*{\thefootnote}{\arabic{footnote}}

The program \norm performs many additional functions, but
was set to perform only vertex enumeration/facet enumeration for these tests. 
By default, it begins with 64-bit integer arithmetic
and only switches to GMP arithmetic (used by all other programs except \porta) in case
of overflow. In this case, all work done with 64-bit arithmetic is discarded.
For our test problems this happens on {\em c30-15}, {\em c40-20} and {\em mit71},
however the first two problems terminated abnormally after switching to GMP.
Using the -B flag \norm will do all computations using GMP arithmetic. We
give times for the default hybrid arithmetic and also for GMP-only arithmetic.
Note that {\em mit71} runs significantly faster with the -B flag reflecting
the time wasted in 64-bit arithmetic mode.

As mentioned above, \porta supports arithmetic using either 64-bit integers or
its own extended precision arithmetic package.
The program terminates if overflow occurs.
We tested both options on each problem and found the extended precision
option outperformed the 64-bit option in all cases.

It is hard to draw many general conclusions from the results in Table \ref{tab:single};
especially since
the four double description implementations behaved remarkably differently 
on most of the problems.
This could be due to the fact that this method is highly sensitive to the insertion order
of the input and the codes may be using different orderings.
One clear result was that none of these codes could solve the cyclic polytope {\em c40-20} problem and
struggled even on {\em c30-15}.
We also observed that the double description codes use substantial memory, especially 
\norm. In fact the machines with 32GB or less of memory 
were not able to solve either {\em mit71} or {\em cp6} using these codes, and even in 
single processor
mode most of the 128GB available on \mait was required for some problems.
Memory use by \lrs/\plrs/\mplrs was negligible, making them good background processes.
On the extremely degenerate problem {\em cp6},
\lrs was in the middle of the pack, about 20\% slower than \cdd, 
whereas \norm was nearly 13 times faster than \lrs.
On this problem, neither \ppl nor \porta was able
to produce any output in the time allotted (76 and 57 days respectively).
Only \porta and \normaliz could effectively solve the sparse 0,$\pm1$ polytope
{\em zfw91}. A 289-core run with \mplrsa was approximately 70 times slower than \porta
and 11 times slower than \normaliz.
Note that with about 151 million cobases/vertex {\em cp6} is far
more degenerate than {\em zfw91}, which has about 4 million cobases/vertex.

To put the above results in perspective, we recall that the problem {\em mit} was a big challenge
in the early 1990s. At that time, early versions of both \cdd and \lrs took over a month
to solve this problem. Combined hardware and software improvements over the years give
speedups of over 5000 times; both codes now complete the job in less than 10 minutes. 
We will see that parallelization of \lrs can lead to further dramatic reductions in running time:
on our 312-core cluster the problem now requires only 12 seconds.

We move now to the three parallel codes.
For \mplrs and \plrs we used the default settings (see User's guide \cite{lrs} for details):
\begin{itemize}
\item \plrs: -id 4
\item \mplrs: -id 2, -lmin 3 -maxc 50 -scale 100 -maxbuf 500
\end{itemize}
For \norm we used the default settings which imply that hybrid arithemtic is used.
Table \ref{tab:su12} contains results for low scale parallelization and all
problems were
run on the single workstation \mait. With 4 cores available, \plrs
usually outperforms \mplrs, they give similar performances with 8 cores,
and \mplrs
is usually faster with 12 or more cores.
With 16 cores \mplrs gave quite consistent speedups, in the range 10-12.3. 
On the problems it could solve, the speedups obtained by
\norm show a much higher variance, in the range 0.93-15.7.

\begin{table}[h]
\centering
\scalebox{0.75}{
\begin{tabular}[ht]{|c||c|c|c||c|c|c||c|c|c||c|c|c||c|c||}
  \hline
  Name& \multicolumn{3}{|c||}{4 cores} & \multicolumn{3}{|c||}{8 cores}
& \multicolumn{3}{|c||}{12 cores} & \multicolumn{3}{|c||}{16 cores}
 \\
    &\multicolumn{3}{|c||}{secs/speedup} & \multicolumn{3}{|c||}{secs/speedup}
& \multicolumn{3}{|c||}{secs/speedup} & \multicolumn{3}{|c||}{secs/speedup}
 \\
       & \mplrs & \plrs & \norm    & \mplrs   &  \plrs & \norm & \mplrs & \plrs & \norm
              &\mplrs   & \plrs  & \norm  \\
  \hline
{\em bv7} &5219 & 2399 & 43 & 1739 & 1213& 23 & 1045 & 818 &17 & 747  &  624  & 13  \\
  &1.6 & 3.5& 1.4 & 4.8 & 6.9& 2.6 & 8.0 & 10.2 & 3.7   & 11.1  & 13.3  & 4.8  \\
  \hline
{\em c30-15}  & 28& 17& ** & 9 & 11&  ** &6  & 9 & **& 4  & 10  & **   \\
 &1.4 & 2.4& -  & 4.4 & 3.6& - & 6.7 & 4.4& -            & 10  & 4  & -  \\
  \hline
{\em c40-20} &5979 &3628& **  &  2023 & 2564&  ** & 1219  & 2237&  **    & 873  & 2066   & **   \\
 & 1.6 &2.6& - & 4.7& 3.7 &- & 7.7  & 4.2 & - & 10.8  & 4.6  & -  \\
  \hline
{\em fq48-19} &146 & 99 &18  & 49 & 52 & 13& 30 &36&12& 21  & 29  & 12   \\
 &1.7 & 2.6&2.3 & 5.2 & 4.7&3.2 & 8.3 & 7.0& 3.4   & 12.0  &12.7   & 3.4  \\
  \hline
{\em mit71}  &11386 & 6479 & 107482 & 3983 & 3320&65507 & 2390& 2254&50910   & 1709  & 1724  & 42916  \\
 &1.8 & 3.2& 4.7   & 5.2&6.2& 7.7   & 8.6& 9.2& 9.9  & 12.1  & 12.0  & 11.7  \\
  \hline
 {\em mit}&293 & 152& 70  & 99 & 89& 42 &61  & 68& 33     & 44  & 57  & 29  \\
 &1.7 &3.3&2.5 &  5.0  &5.6 & 4.2  &8.1  &7.3 & 5.3   &  11.3 & 8.7  & 6  \\
  \hline
{\em perm10} &1422 &709& 1085  & 481 & 445& 960 &292  & 367 &1090     & 215  & 320  & 1093  \\
 &1.6 & 3.2& .94&4.7 &5.0& 1.1   &7.7& 6.1& .94   & 10.4  & 7.0  & .93  \\
  \hline
{\em zfw91} &* &*& 46741  & * & *& 23885 &*  &*  & 15975    & *  & *  & 12110   \\
 &-&-& 4.1&-&-& 7.9   &-&- & 11.9   & - & - & 15.7  \\
  \hline
\hline
{\em cp6} &968550  &486667 & 42774   &331235 &268066& 23493 &199501 & 201792& 18585     & 143006  & 169352  & 16980   \\
 & 1.8&3.6& 3.2  & 5.3&6.6& 5.9   &8.8 &8.7& 7.4   & 12.3  & 10.4  & 8.1  \\

\hline

\end{tabular}
}
\caption{Small scale parallelization(\mait):  *=time $>$ 604800 secs, **=abnormal termination}
\label{tab:su12}
\end{table}

Table \ref{tab:su64} contains results for medium scale parallelization on
the 64-core shared memory machine \mais. Note that these processors are considerably
slower than \mait on a per-core basis.
We used 8,16,32,64 cores and speedups are
measured by comparing with the running time on 8 cores. 
With 64 cores,
\mplrs was the clear winner over \plrs with speedups ranging
from 4.3 to 7.2.
\plrs showed little improvement after 32 cores and \norm again had very large variance.

\begin{table}[htbp]
\centering
\scalebox{0.75}{
\begin{tabular}[ht]{|c||c|c|c||c|c|c||c|c|c||c|c|c||c|c|c|}
  \hline
  Name& \multicolumn{3}{|c||}{8 cores} &\multicolumn{3}{|c||}{16 cores} & \multicolumn{3}{|c||}{32 cores}
& \multicolumn{3}{|c||}{64 cores}
 \\
    &\multicolumn{3}{|c||}{secs/speedup vs 8 cores}  &\multicolumn{3}{|c||}{secs/speedup vs 8 cores} & \multicolumn{3}{|c||}{secs/speedup vs 8 cores}
& \multicolumn{3}{|c||}{secs/speedup vs 8 cores}
 \\
     & \mplrs & \plrs & \norm   & \mplrs & \plrs & \norm    & \mplrs   &  \plrs & \norm & \mplrs & \plrs & \norm
              \\
  \hline
{\em bv7}   & 3238 & 2255  & 60 &1478 & 1212 & 39 & 1206 & 726& 29 &515  & 506 &21 \\
   & 1 & 1  & 1  &2.2 &1.9 &1.5  &2.7  &3.1 & 2.1 &6.3  &4.4  &2.9    \\
\hline
{\em c30-15}   & 17  & 22  & ** &9 & 20& **  & 5 & 22& ** & 4  & 21 & ** \\
   & 1 & 1  & -  &1.9 &1.1 & - & 3.4 &1.0 & - & 4.3  & 1.0 & -    \\
\hline
{\em c40-20}   & 3882  & 4694 & ** &1876 &4163&  **  & 1141  & 4192& **  &717   & 4086& **    \\
   & 1 & 1  & -  &2.1 &1.1 & - & 3.4 &1.1 & - &  5.4 & 1.1  &  -  \\
\hline
{\em fq48-19}   & 89  & 95  & 28 &42 & 57& 24 & 23 & 39 &22 &14  &31&23\\
   & 1 & 1  & 1  & 2.1 &1.7 & 1.2&3.9  &2.4 & 1.3 &6.4  &3.1  & 1.2    \\
\hline
{\em mit71}   & 7395 & 6218  & 115088 &3401 & 3441 &77436 &1900  & 2130& 60694&1251  & 1640& 51594  \\
   & 1 & 1  & 1  & 2.2 & 1.8 & 1.5 & 3.9 & 2.9 & 1.9  & 5.9  & 3.8  & 2.2    \\
\hline
 {\em mit}   & 195  & 175 & 111 & 93& 123& 83   & 53 & 120 & 75 &42 & 124&  82    \\
   & 1 & 1  & 1  & 2.1 & 1.4 & 1.3 & 3.7  & 1.5& 1.5  &4.6  &1.4  & 1.4    \\
\hline
{\em perm10}   & 909  & 841   & 1951&432 &617&1870   & 253 & 569& 1840 & 171 & 573 & 1930    \\
   & 1 & 1  & 1  & 2.1 & 1.4 & 1.1 & 3.6 & 1.5 & 1.1  & 5.3  & 1.5  & 1   \\
\hline
{\em zfw91}   & *  & *   & 42409 &* &* &24822  & * &* &14452  & * & * & 7332    \\
   & - & -  & 1  & -& -& 1.7 & -& -&2.9  & -   & - & 5.8   \\
\hline
\hline
{\em cp6 \tablefootnote{\plrs times computed using v6.0}}   &727771   & 565915  & 38621 & 326214 &377857& 23773  & 171194  &298408& 17468 & 100676  &229713 &  15480     \\
 & 1 & 1  & 1 &2.2 & 1.5 &1.6  & 4.3  & 1.9 &2.2  & 7.2  &2.5  & 2.5    \\
\hline

\end{tabular}
}
\caption{Medium scale parallelization (\mais):  *=time $>$ 604800 secs, **=abnormal termination}
\label{tab:su64}
\end{table}

Table \ref{tab:su96} contains results for medium scale parallelization
on a 312-core cluster of computers. 
Only \mplrs is able to use all cores in this heterogeneous environment.
The machines were scheduled in the order given at the end of Section \ref{alg} (excluding \tsubame).
Due to the heterogeneous selection of machines we do not present speedups in this table.
For example, we observed that \mait is substantially faster than the other machines --
more than would be expected by simply comparing clock speeds and number of cores.
It was more than
twice as fast as \mai on {\em c40-20}.
Jobs completing in under a minute do not profit much, if at all, as extra cores are added.
However, the longer running jobs show continuous improvement.
Excluding {\em zfw91},
\lrs required about 3 weeks on the fastest machine (\mait) to complete 
the other 9 problems.
Using the 312-core cluster this time is reduced to 4 hours and 40 minutes.
These total times are dominated by {\em cp6}. Excluding this problem as well, the \lrs
total running time of 12 hours 13 minutes is improved to roughly 8 minutes using the cluster.

\begin{table}[htbp]
\centering
\scalebox{0.9}{
\begin{tabular}[ht]{|c||c|c|c|c|c|c|}
  \hline
 Name& \multicolumn{6}{|c|}{\mplrs}  \\
  & 16 cores  & 32 cores
& 64 cores  & 128 cores & 256 cores & 312 cores
 \\
  & secs  &secs & secs
& secs & secs & secs
 \\
  \hline
{\em bv7} & 747 &389 &262  &179 & 101 & 88 \\
  \hline
{\em c30-15}   &4 &3& 2 & 3 & 2 & 2 \\
  \hline
{\em c40-20}  &873 &528 &328   &218 & 133 & 121\\
  \hline
{\em fq48-19} & 21 & 11 & 7 & 5 & 4 & 5   \\
  \hline
{\em mit71} &  1709 & 956 & 625& 421 & 228 & 199   \\
  \hline
{\em mit}  &44 &26   &21    & 23 & 13 &  12 \\
  \hline
{\em perm10}  &215 & 118 & 89& 75 & 53 & 55 \\
  \hline
  \hline
{\em cp6} & 143006 &  75712 & 50225  & 33684   & 18657   & 16280   \\
  \hline
\end{tabular}
}
\caption{Medium scale parallelization(cluster)}
\label{tab:su96}
\end{table}

Table \ref{tab:su1200} shows results for large scale parallelization
obtained by Kazuki Yoshizoe using \mplrs v.\ 6.0 on the \tsubame supercomputer at
the Tokyo Institute of Technology. He ran tests using problems {\em mit71} and {\em cp6}
and observed near linear speedup between
12 and 1200 cores for both problems\footnote{{\em cp6} benchmark was taken with \mai
which has a similar processor (Xeon X5650) to those we used on \tsubame (Xeon X5670).}.
With 1200 cores \mplrs~solved {\em cp6}
in about 42 minutes, nearly 600 times faster than \cdd, 55 times
faster than \norm in single processor mode
and over 6 times faster than \norm running on 64 cores, the largest shared memory machine available to us. 

\begin{table}[htbp]
\centering
\scalebox{0.9}{
\begin{tabular}[ht]{|c||c|c|c|c|c|c|c|c|c|c|}
  \hline
  Name& \multicolumn{7}{|c|}{\mplrs (v.\ 5.1b)}  \\

    & 12 cores & 36 cores & 72 cores & 144 cores & 300 cores & 600 cores & 1200 cores \\
  \hline
{\em cp6} &283403(\mai) & * & * &20383  & 9782  & 4913  & 2487     \\
   &  1 & -  & - & 14 & 29 & 58   & 114   \\
  \hline
{\em mit71}  &4207 & 1227 & 602 & 297  & 146& 81 & 45     \\
   & 1  & 3.4   & 7.0   & 14  & 29   & 52   & 94   \\
  \hline
\end{tabular}
}
\caption{Large scale parallelization: secs/speedups, *=\tsubame~time limit exceeded}
\label{tab:su1200}
\end{table}

\section{Conclusions}
These results show that the difficulty of solving vertex/facet enumeration problems
varies enormously, even for inputs of roughly the same size. Any given problem
may be tractable or intractable depending on the method used to solve it. General rules are dangerous
and likely to be contradicted by further examples, but we hazard two:
learn about your polytope and use multicore hardware.

\subsection{Learn something about your polytope}
Unfortunately not much can be learned by simply inspecting the input file. Many 0/1 input files
are highly degenerate, but not all: {\em perm10}, for example, is a simple polytope.
Fortunately the degeneracy of a polytope can be checked by doing a partial run of \lrs for a few minutes, stopping
after a certain number of bases have been computed. As seen from Table \ref{polytopes}, the ratio of bases computed
to V/H output gives a good estimate of degeneracy of the problem. It will also give an indication
as to whether the output is binary ({\em cp6}), consists of small integers ({\em bv7}, 
{\em perm10}, {\em zfw91}), 
huge integers ({\em c30-15}, {\em c40-20}) or rationals ({\em fq48-19}, {\em mit}, {\em mit71}). 
\lrs also has an estimate feature that gives an unbiased estimate of the output size, number of bases and
total \lrs running time. These estimates have high variance but do give some indication of the
tractability of the problem.

For problems with low degeneracy or very large output sizes pivoting methods such as the \lrs
family may be the only tractable approach. For extremely degenerate problems with binary or
small integer output it is not so clear, as can be seen by comparing the results obtained for {\em cp6} and
{\em zfw91}.

\subsection{Use multicore hardware}
Comparing Table \ref{tab:single} with the remaining tables clearly indicates the necessity of using
parallel processing for hard vertex/facet enumeration problems: even just 16 cores gives an order of
magnitude improvement.
A supercomputer on the scale of \tsubame may seem out of reach for most researchers. However, at current prices,
a 1200-core cluster could be built for roughly \$100,000 and would be considerably cheaper with used hardware.
This price will certainly fall substantially in the near future making this amount of computing power
readily available to more researchers. The problem will not be the availability of the hardware but
the availability of software that can make effective use of it.

\vspace{-0.08in}
\section*{Acknowledgements}
We thank Kazuki Yoshizoe for kindly allowing us to use the results of his
\tsubame experiments and for helpful discussions concerning the MPI library
which improved \mplrs' performance. This work was partially supported by JSPS Kakenhi Grants 23700019 and 15H00847,
Grant-in-Aid for Scientific Research on Innovative Areas, `Exploring
the Limits of Computation (ELC)'.
\vspace{-0.08in}
\bibliographystyle{spmpsci}
\bibliography{mplrs}

\end{document}